\pdfoutput=1
\documentclass[12pt,preprint]{aastex}
\usepackage{graphicx}
\usepackage{natbib}
\usepackage{aas_macros}
\usepackage[section] {placeins}
\usepackage{subfigure}
\usepackage{float}
\usepackage{color}

\usepackage[scaled]{helvet}

\usepackage[T1]{fontenc}

\def\msun{{\rm\,M_\odot}}

\def\msun{{\rm\,M_\odot}}

\newcommand{\kms}{\, {\rm km\, s}^{-1}}

\newcommand{\mpc}{\, {\rm Mpc}}
\newcommand{\kpc}{\, {\rm kpc}}
\newcommand{\hmpc}{\, h^{-1} \mpc}

\newcommand{\hi}{\mbox{H\,{\scriptsize I}\ }}
\newcommand{\heii}{\mbox{He\,{\scriptsize II}\ }}

\def\h2{${\rm\,H_2}$}

\makeatletter 

\def\hi{\noindent \hangindent=2.5em}

\def\kpc{{\rm\,kpc}}

\def\mpc{{\rm\,Mpc}}

\def\kms{{\rm\,km/s}}
\def\msun{{\rm\,M_\odot}}

\def\vol#1  {{{#1}{\rm,}\ }}

\def\hi{{H\thinspace I}\ }

\def\heii{{He\thinspace II}\ }

\newcount\refno
\newcount\rfno
\def\eq{$^{\the\refno\ }$\advance\refno by 1}
\def\ad{\advance\rfno by 1}

\def\clock{\count0=\time \divide\count0 by 60
     \count1=\count0 \multiply\count1 by -60 \advance\count1 by \time
     \number\count0:\ifnum\count1<10{0\number\count1}\else\number\count1\fi}

\def\myputfigure#1#2#3#4#5%
{\hskip0.03\textwidth\vskip#5pt
\makebox[0pt]{\hskip#2in
\includegraphics[width=#3\textwidth]{#1}}\vskip#4pt\hfill}

\newcount\refno
\newcount\rfno
\def\eq{$^{\the\refno\ }$\advance\refno by 1}
\def\ad{\advance\rfno by 1}

\definecolor{burntorange}{rgb}{1,0.4,0.2}

\definecolor{burntorange}{rgb}{1,0.4,0.2}

\def\mbh{{\rm\,M_{BH}}}
\def\mbg{{\rm\,M_{BG}}}

\def\mnsc{{\rm\,M_{NSC}}}

\usepackage{color}

\begin{document}

\title{Coevolution Between Supermassive Black Holes and Bulges Is Not Via Internal Feedback Regulation But By Rationed Gas Supply 
Due To Angular Momentum Distribution}

\author{Renyue Cen\altaffilmark{1}}

\footnotetext[1]{Princeton University Observatory, Princeton, NJ 08544;
 cen@astro.princeton.edu}

\begin{abstract} 

We reason that, without physical fine-tuning, neither the supermassive black holes (SMBHs)
nor the stellar bulges can self-regulate or inter-regulate by driving away already fallen cold gas to 
produce the observed correlation between them.
We suggest an alternative scenario where the observed mass ratios of the SMBHs to bulges
reflect the angular momentum distribution of infallen gas such
that the mass reaching the stable accretion disc 
is a small fraction of that reaching the bulge region, averaged over the cosmological time scales.
We test this scenario using high resolution, large-scale 
cosmological hydrodynamic simulations (without AGN feedback), 
assuming the angular momentum distribution of gas landing in the bulge region to yield a Mestel disc
that is supported by independent simulations resolving the Bondi radii of SMBHs.
A mass ratio of $0.1-0.3\%$ between the very low angular momentum gas that free-falls 
to the sub-parsec region to accrete to the SMBH and the overall star formation rate is found.
This ratio is found to increase with increasing redshift to within a factor of $\sim 2$,
suggesting that the SMBH to bulge ratio is nearly redshift independent, 
with a modest increase with redshift, a testable prediction.
Furthermore, the duty cycle of active galactic nuclei (AGN) with high Eddington ratios
is expected to increase significantly with redshift.
Finally, while SMBHs and bulges are found to coevolve on $\sim 30-150$Myr time scales or longer,
there is indication that, on shorer time scales, the SMBH accretion rate and star formation may be 
less correlated.

\end{abstract}

\section{Introduction}

There is mounting evidence that massive bulges in the nearby universe
harbor central SMBHs of mass $10^6-10^9\msun$. 
The correlation between SMBH mass ($\mbh$) and the bulge (BG) mass ($\mbg$) or velocity dispersion ($\sigma$) 
\citep[e.g.,][]{1998Magorrian,1998Richstone,2000Gebhardt,2000Ferrarese,2002Tremaine}
suggests coevolution. 
Although alternative models for producing this observed relation are available
\citep[e.g.,][]{2000Ostriker, 2001Adams, 2003Colgate, 2007Cen},
the correlation is often construed as evidence for AGN feedback to regulate the growth of SMBHs and bulges.
The idea that AGN feedback may alleviate problems in galaxy formation models
\citep[e.g.,][]{2000Kauffmann,2006Croton,2008Somerville}
further enhances its appeal.
The three-dimensional hydrodynamic simulations  
successfully reproduced the observed $\mbh$/$\mbg$ ratio \citep[e.g.,][]{2005DiMatter,2006Hopkins},
providing the physical basis for this scenario.

This {\it Letter} has two goals.
First, we make a qualitative examination of the implications of the observed relation between bulges
and the central massive objects (CMOs),
wherein the two follow a linear relation over four decades in mass.
It is shown that
neither the SMBHs nor the nuclear star clusters (NSCs) nor the stellar bulges 
could have played a dominant role in regulating the growth of any of the three components in the way
of blowing away a significant fraction of gas already landed in the respective regions so as to produce the CMO-bulge relation.
Second, an alternative model is put forth wherein the correlation between SMBH mass and bulge mass
is 
dictated by the angular momentum distribution of the infalling gas.
We successfully test this new scenario 
using {\it ab initio}
{\color{red}\bf L}arge-scale {\color{red}\bf A}daptive-mesh-refinement {\color{red}\bf O}mniscient {\color{red}\bf Z}oom-{\color{red}\bf I}n ({\color{red}\bf LAOZI}) cosmological hydrodynamic simulations. 

\section{Arguments Against Internal Regulation of the Central Components}

With the ACS Virgo Cluster Survey of  
early-type galaxies spanning four decades in mass, 
\citet[][]{2006Cote} and \citet[][]{2006Ferrarese} find 
a transition at $M_{B,0}=-20.5$, 
where the brighter galaxies lack resolved stellar nuclei and SMBHs dominate the CMO mass, 
while fainter ones have resolved stellar nuclei that dominate the CMO mass.
Furthermore, the logarithm of the mean nucleus-to-galaxy luminosity ratio in fainter, nucleated galaxies, 
$-2.49\pm -0.09$ ($\sigma =0.59\pm -0.10$) is indistinguishable from that of the 
SMBH-to-bulge mass ratio, 
$-2.61\pm -0.07$ ($\sigma =0.45\pm -0.09$). 
A similar result is found by \citet[][]{2006Wehner} using a different data set. 
\citet[][]{2012Turner} 
find an identical relation 
using early-type galaxies 
in the ACS Fornax Cluster Survey.
We express the universal scaling relation between CMOs and bulges as
\begin{equation}
\label{eq:mass}
{\rm M_{CMO}} = \mbh + \mnsc = \alpha \mbg,
\end{equation}
\noindent
whereby 
with the transition between NSC and SMBH occurs at $M_B\sim -20.5$ or stellar mass $\mbg_{0}=(3-4)\times 10^{10}\msun$,
and $\alpha = 2.5\times 10^{-3}$
\citep[][]{2006Cote, 2006Ferrarese, 2006Wehner, 2012Turner}.
One may express regulation of the growth of bulges as
\begin{equation}
\label{eq:energyblg}
{\rm e_{BH} \mbh + e_{NSC} \mnsc + e_{BG} \mbg = f \sigma^\beta\mbg},
\end{equation}
\noindent
where ${\rm e_{BH}}$, ${\rm e_{NSC}}$ and ${\rm e_{BG}}$ are the feedback strength coefficients per unit mass
of the respective components exerted on the stellar bulge
and the ejected gas mass is equal to ${\rm fM_{BG}}$;
$\sigma$ is the velocity dispersion of the stellar bulge;
$\beta$ is a parameter that absorbs uncertainties regarding the dynamics of concerned feedback processes,
with $\beta=2$ for energy-conserving feedback (${\rm e_{BH}}$,  ${\rm e_{NSC}}$ and ${\rm e_{BG}}$ have units of energy per unit mass) 
 and $\beta=1$ for momentum-conserving feedback 
(${\rm e_{BH}}$,  ${\rm e_{NSC}}$ and ${\rm e_{BG}}$ have units of momentum per unit mass).
Note that a significant feedback regulation means $f\gg 1$.

Insights can be gained by asking the following question: 
Can the feedback from SMBH and NSCs conspire to regulate the growth of the stellar bulge,
i.e.,
\begin{equation}
\label{eq:cmo}
{\rm e_{BH} \mbh + e_{NSC} \mnsc = f \sigma^\beta\mbg}?
\end{equation}
\noindent
The single powerlaw relation between ${\rm M_{CMO}}$ and $\mbg$ across four decades in bulge mass 
can be understood, only if the negative feedback per unit stellar mass of the NSC 
and of the SMBH are approximately the same,
${\rm e_{BH}\approx e_{NSC}}$, barring the unknown physical reason for 
the right hand side of Eq~(\ref{eq:cmo}) 
- the required amount of notional feedback to regulate the bulge growth - 
to change character abruptly at $\mbg = \mbg_{0}$.

Although having ${\rm e_{BH}\approx e_{NSC}}$ may be possible,
it would render a negative answer to the question above (Eq~\ref{eq:cmo}), 
as follows.
In the momentum driven regime, since the feedback from the nuclear cluster 
is subject to higher densities and shorter cooling timescales 
hence diminished strength in comparison to that in the stellar bulge, i.e., ${\rm e_{BG} > e_{NSC}}$.
In the energy driven feedback scenario, ${\rm e_{BG} = e_{NSC}}$. 
Since ${\rm M_{NSC}\ll \mbg}$, the supernova feedback from stars in the bulge 
would vastly exceed that from the NSC.  
This thus invalidates the statement that
the NSC and SMBH provide the necessary feedback to regulate the growth of the bulge.

The only scenario left for the SMBH to regulate the bulge growth is to 
force ${\rm e_{NSC}=0}$ and assume the feedback per unit SMBH mass, while constant at $\mbg > \mbg_{0}$,
to become negligible at about $\mbg = \mbg_{0}$.
In both the momentum \citep[$\beta=1$,][]{2010Ostriker} 
and energy feedback scenario \citep[$\beta=2$,][]{2012FG},
the amount of momentum or energy per unit SMBH mass, ${\rm e_{BH}}$, 
is ultimately proportional to the driving energy ($\propto \mbh c^2$, where $c$ is speed of light).
Thus, there exists no known process to suddenly make ${\rm e_{BH}}$ drop to zero at some specific $\mbh$,
while being constant otherwise.

If negative feedback is needed to internally regulate the bulge, the only alternative left is stellar feedback from bulge stars 
themselves, i.e.,
\begin{equation}
\label{eq:blg}
{\rm e_{BG} = f \sigma^\beta}.
\end{equation}
\noindent
Under the assumption that the feedback strength from stars per unit mass (${\rm e_{BG}}$) 
is constant, 
one obtains $f \propto \sigma^{-\beta}$,
which has the same dependence on $\sigma$ as the predicted mass loading factors
for both momentum ($\beta=1$) or energy ($\beta=2$) driven winds \citep[e.g.,][]{2005Murray}.
Therefore, bulge self-regulation, if required, would be physically supportable and self-consistent.
If bulge is self-regulated, then, under the assumption that ${\rm e_{NSC} = e_{BG}}$,  NSC may also be self-regulated.
The correlation between ${\rm M_{MCO}}$ and $\mbg$ would then require that the mass loading factor 
for the SMBH is the same as for the NSC, i.e.,  ${\rm e_{BH} = e_{NSC}}$, which is a fine-tuned outcome.
In the absence of inter-regulation between CMOs and bulges, 
the proportions of the amount of gas feeding the nuclear and bulge regions 
must be proportional to the observed ${\rm M_{CMO}}/\mbg$ ratio. 


\section{An Alternative Scenario: Rationed Cold Gas Supply to Nuclear and Bulge Regions Over Cosmological Time Scales}\label{sec: prop}

Our arguments in the previous section indicate that 
the observed ${\rm M_{CMO}}$-$\mbg$ correlation 
requires the same proportionality in the initial amounts of gas feeding the respective regions,
averaged over the cosmological time scales.
We test this scenario using direct cosmological simulations.

\subsection{Simulation Characteristics}

See \citet[][]{2014Cen} for a more detailed description
of the {\it ab initio} {\color{red}\bf LAOZI} simulations.
Briefly, we use 
the WMAP7-normalized \citep[][]{2011Komatsu} $\Lambda$CDM model:
$\Omega_M=0.28$, $\Omega_b=0.046$, $\Omega_{\Lambda}=0.72$, $\sigma_8=0.82$,
$H_0=100 h \,{\rm km\, s}^{-1} {\rm Mpc}^{-1} = 70 \,{\rm km\, s}^{-1} {\rm Mpc}^{-1}$ and $n=0.96$.
A zoom-in box of size $21\times 24\times 20h^{-3}$Mpc$^3$ comoving is embedded in a $120~h^{-1}$Mpc periodic box.
The maximum resolution is better than $111h^{-1}$pc (physical) at all times.
Star formation follows the prescription of \citet[][]{1992CenOstriker}.
Supernova feedback from star formation is modeled following \citet[][]{2005Cen}
with feedback energy being distributed into 27 local gas cells 
weighted by the specific volume of each cell,
to 
mimic the process of supernova blastwave propagation to channel more energy into the less dense regions.
We exclude AGN feedback in order to ascertain the lack of need for it.

\subsection{Construction of Gas Feeding Histories of Simulated Galaxies}
\label{ssec:mergertrees}

Galaxies 
are identified using the HOP algorithm 
\citep[][]{1998Eisenstein} grouping stellar particles. 
Galaxy catalogs are constructed from $z=0.62$ to $z=1.40$ with an increment of $\Delta z=0.02$ and 
from $z=1.40$ to $z=6$ with $\Delta z=0.05$, 
having a temporal resolution of $30-150$Myr.
For each galaxy at $z=0.62$ a genealogical line is constructed 
up to $z=6$, where the parent of each galaxy is identified with the one at the next higher redshift
with the most overlap in stellar mass.
At each redshift, we compute the amount (${\rm M_{c}}$) and mean specific angular momentum (${\rm J_{c}}$) of gas
in the central $1$kpc region.
To proceed, an ansatz is made: the gas mass with angular momentum lower than ${\rm J_{n}}$ is 
${\rm M_{c}(\beta J_{n}/(1+\beta)J_{c}})^\beta$.
We use $\beta=1$, 
which corresponds to a \citet[][]{1963Mestel} disc of surface density $\Sigma(r)\propto r^{-1}$. 
$\beta=1$ is motivated by simulations of \citet[][]{2010Hopkins, 2011Hopkins} with resolution as high as 0.1pc. 
Figure 12 of \citet[][]{2010Hopkins}
shows that the evolved density runs of the gas discs, on average, follow the $\Sigma(r) \propto r^{-1}$ profile from 0.1pc to 1kpc. 
In all of the six individual cases with significant gas inflow, 
shown in Figures (2, 3) of \citet[][]{2011Hopkins}, the $\Sigma(r) \propto r^{-1}$ profile 
provides an excellent fit.  
We compute the 
1-d stellar velocity dispersion $\sigma$ within the effective radius 
for each galaxy in the simulation at any redshift 
and assume an SMBH of mass equal to $\mbh=10^8\msun (\sigma/200\kms)^4$ \citep[][]{2002Tremaine}.
The Bondi radius is 
\begin{equation}
\label{eq:rB}
{\rm r_B = 2G\mbh/3\sigma^2 = 7.2pc (\sigma/200\kms)^2},
\end{equation}
\noindent
and the specific angular momentum at ${\rm r_B}$ is 
\begin{equation}
\label{eq:JB}
{\rm J_B=\sqrt{2}r_B\sigma}. 
\end{equation}
\noindent
The gas landing within ${\rm r_0}$ is assumed to accrete to the SMBH, where at ${\rm r>r_0}$ 
the disc has Toomre $Q$ parameter below unity and is hence consumed by star formation.
Expressing various parameters by their fiducial values, we have 
\begin{equation}
\label{eq:r0}
{\rm r_0= 0.42(\alpha/0.1)^{2/5}(l_{E}/0.1)^{-2/5}(\mbh/10^8\msun)^{3/25}(\mathit{Ma}/0.1)^{14/25}(\kappa/\kappa_e)^{4/25}~pc}
\end{equation}
\noindent
\citep[Eq 42,][]{2003Goodman},
where $\alpha$ is radiative efficiency, $l_E$ luminosity in Eddington units, 
$\mathit{Ma}$ Mach number of the viscous disc at $r_0$,
and 
$\kappa$ and $\kappa_e$ opacity and electron-scattering opacity, respectively.
Hence the feeding rate to the accretion disc that eventually accretes to the SMBH is 
\begin{equation}
\label{eq:Mfeed}
{\rm \dot M_{feed}= M_{c}((r_0/r_B)^{1/2}J_{B}/J_{c})t_{dyn}^{-1}},
\end{equation}
\noindent
where 
the angular momentum at $r_0$ is ${\rm J_0 = (r_0/r_B)^{1/2}J_B}$ for a Keplerian disc
and ${\rm t_{dyn}=1kpc/\sqrt{3}\sigma}$ is the free-fall time at $1$kpc.
For our analysis, we use
\begin{equation}
\label{eq:r0f}
{\rm r_0= 0.42(\mbh/10^8\msun)^{3/25}~pc}, 
\end{equation}
\noindent
bearing in mind that uncertainties are at least on the order of unity.
To see how uncertainty in $\beta$ affects results,
we note, a 25\% deviation in $\beta$ from unity causes
${\rm \dot M_{feed}}$ in Eq~(\ref{eq:Mfeed}) to change by a factor of $2.7$, 
which can be compensated by adjusting
each of the parameters in Eq~(\ref{eq:r0}) except $\mbh$ by a factor of $2.5$ appropriately.

\subsection{Results}

\begin{figure}[!h]
\centering
\vskip -0.0cm
\resizebox{6.5in}{!}{\includegraphics[angle=0]{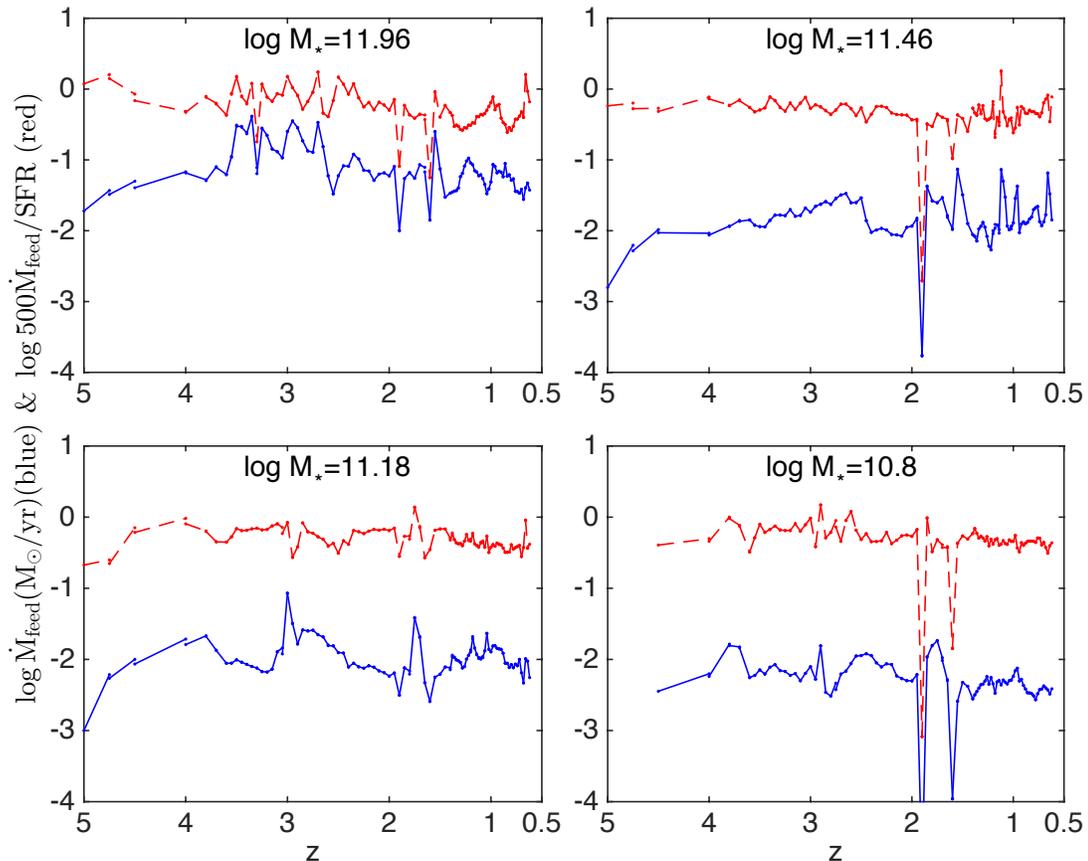}}
\vskip -0.5cm
\caption{
shows histories of the feeding rate ${\rm \dot M_{feed}}$ (blue) 
and ${\rm R\equiv 500\dot M_{feed}/SFR}$ (red) 
for four random galaxies.
The logarithm of the stellar mass for each galaxy at $z=0.62$ is indicated at the top of each panel.
}
\label{fig:Jlow}
\end{figure}

We define a ratio ${\rm R\equiv 500\dot M_{feed}/SFR}$ (${\rm SFR}$ is the star formation rate) 
such that, if ${\rm R}$ is about unity, the observed SMBH to bulge mass ratio of $\sim 0.2\%$ 
\citep[e.g.,][]{2003Marconi,2004Haring} would be borne out.
Transformation from stellar disc(s) to a bulge is not addressed here.
It is noted, however, that stellar discs formed from multiple gas inflows of inclined angles 
over the lifetime of a galaxy may be conducive to bulge formation.
Note that ${\rm SFR}$ is computed directly during the simulation,
whereas the SMBH accretion rate is computed in post-processing by evaluating 
Eq~(\ref{eq:Mfeed}). 
Figure~\ref{fig:Jlow} shows histories of ${\rm \dot M_{feed}}$ (blue) 
and ${\rm R}$ (red) for four random example galaxies.
The most noticeable feature is that, without any intentional tuning, ${\rm R}$
hovers close to unity with fluctuations of order unity.

\begin{figure}[!h]
\centering
\vskip -0.0cm
\resizebox{6.5in}{!}{\includegraphics[angle=0]{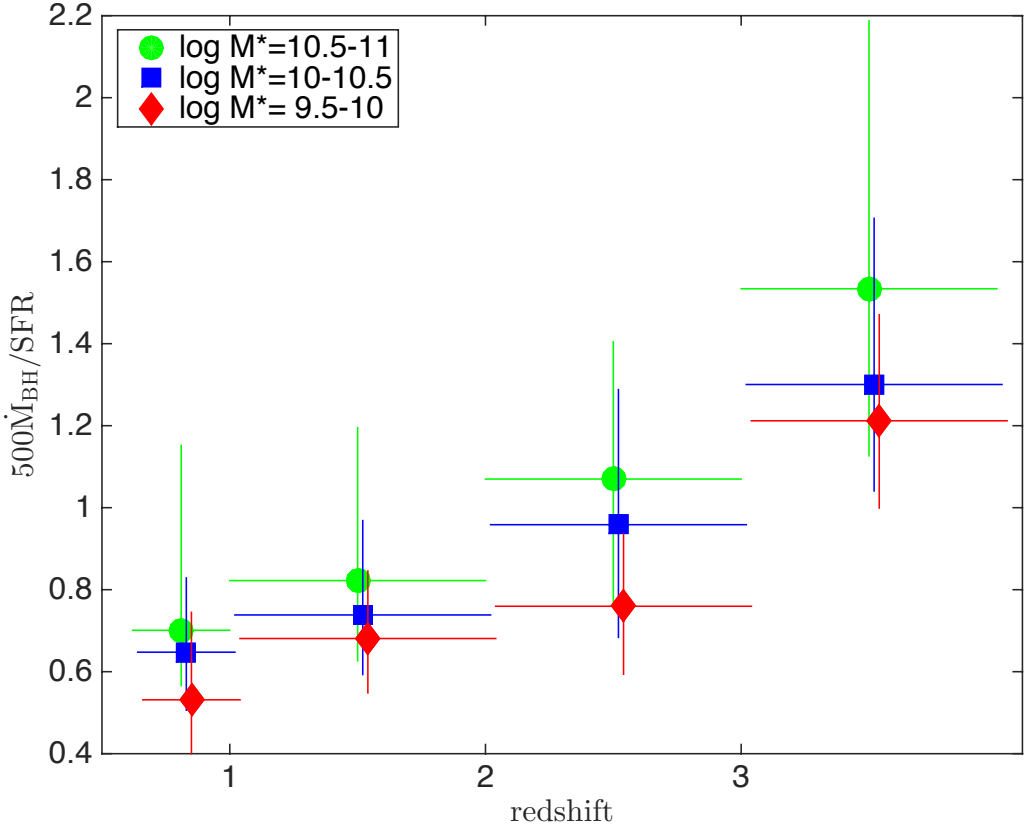}}
\vskip -0.5cm
\caption{
shows the median of ${\rm R}$ as a function of redshift,
separately for three stellar mass ranges 
$10^{9.5-10}\msun$ (red),
$10^{10-10.5}\msun$ (blue)
and
$10^{10.5-11}\msun$ (green).
The stellar mass is measured at the redshift in question.
The vertical errorbars indicate the interquartile range,
whereas the horizontal errorbars represent the redshift range of the bin.
The red and blue points are horizontally slightly right-shifted for clarity of display.
There are $(659, 2214)$ galaxies with stellar mass in the range $10^{10.5-11}\msun$ for $z=(3-4, 0.62-1)$, respectively.
}
\label{fig:ratz}
\end{figure}

Figure~\ref{fig:ratz} shows ${\rm R}$ as a function of redshift. 
We see that ${\rm R}$ increases 
with increasing redshift from $\sim 0.7$ at $z=0.6-1$ to $\sim 1.5$ at $z=3-4$ for galaxies 
with $10^{10.5-11}\msun$ (green), with similar trends for other mass ranges.
We highlight three implications.
First, the observed SMBH to bulge ratio is readily achievable in a cosmological setting,
with a slight tendency for more massive galaxies to have higher ${\rm R}$. 
This is due to the rationing of gas supply to the central regions of galaxies: 
a small amount of gas of the lowest angular momentum feeds the SMBH accretion disc, while the rest builds up the stellar bulge,
with the demarcation line determined by the accretion disc stability condition.
Note that our analysis is solely based on the angular momentum distribution of gas
that has already landed in the central $1$kpc region. 
The frequency of gas inflow events into the central regions and the mass distribution of events
are computed directly in our simulations. 
Second, ${\rm R}$ increases with increasing redshift, to within a factor of $\sim 2$.
The trend with redshift is expected in a cosmological context, 
because both the frequency and strength of galaxy interactions
increase with increasing redshift, yielding overall inflow gas of lower angular momentum hence a larger $R$ at high redshift.
Third, the smoothness of ${\rm R}$ on cosmological time scales ($\ge 100$Myr)
suggests that the dispersion of ${\rm R}$ is modest, around order unity, at all redshifts,
consistent with the dispersion of the observed correlation locally 
(note that the comparison is made between computed  $\dot M_{\rm feed}/{\rm SFR}$ and observed $\mbh/\mbg$).
Future observations at high redshift may be able to test these predictions.
Although $R$ is relatively smooth over cosmological time scales,
the gas inflow rate varies up to an order of magnitude (Figure~\ref{fig:Jlow}).
The fluctuations in the inflow rate are caused by a variety of physical processes, including interactions between galaxies in close proximity,
minor mergers and occasional major mergers. 
We have not studied in sufficient detail to ascertain whether secular processes play any major role.

\begin{figure}[!h]
\centering
\vskip -0.0cm
\resizebox{6.5in}{!}{\includegraphics[angle=0]{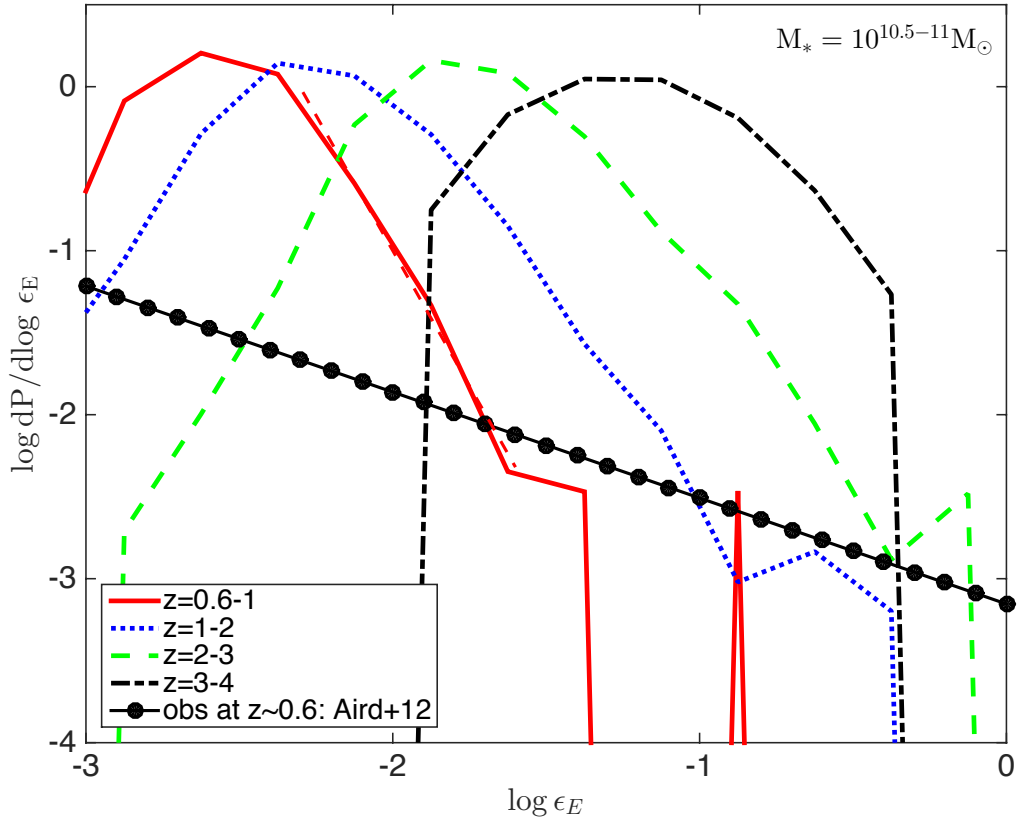}}
\vskip -0.5cm
\caption{
shows the probability distribution of feeding rate in units of Eddington rate per logarithmic Eddington ratio interval, 
as a function of Eddington ratio,
in four redshift ranges, $z=0.62-1$ (solid red), $z=1-2$ (dotted blue), $z=2-3$ (dashed green),
and $z=3-4$ (dot-dashed black) for galaxies in the stellar mass range of $10^{10.5-11}\msun$
(other stellar mass ranges have similar properties).
Also show as solid dots is the observed powerlaw distribution with a slope of $\sim -0.6$ at $z\sim 0.6$ from \citet[][]{2012Aird}.
The slope of the solid red curve is $-3.3$ measured for the $\log {\rm e_E}$ range from $-2.3$ to $-1.6$ indicated
by the red dashed line.
}
\label{fig:duty}
\end{figure}

Is SMBH accretion rate directly dictated by the feeding rate from galactic scales?
Figure~\ref{fig:duty} shows 
the probability distribution of feeding rate in units of Eddington rate as a function of Eddington ratio.
The Eddington ratio is based on the assumed $\mbh$ from the observed $\mbh-\sigma$ relation. 
At $z\sim 0.6$ where comparisons with observations may be made, 
the computed distribution is steeper, computed slope $-3.3$ versus $-0.60$ observed.
This indicates that accretion onto the SMBHs is ``filtered" through physical processes operating on the accretion disc.  
This suggests that temporal correlation between AGN and star formation activities in individual galaxies
below $30-150$Myr is expected to be weak, in excellent agreement with observations \citep[e.g.,][]{2014Hickox}.
A comparison between the distribution of the feeding rate to the accretion disc (red curve)
and that of the observed Eddington ratio (black dots)
suggests that at $z\sim 0.6$ accretion discs around SMBHs spend most of the time accumulating gas,
at feeding rate below $1\%$ Eddington ratio
and that the apparent powerlaw distribution of Eddington ratio may be a result of 
superposition of AGN internal light profiles that are universal in shape (i.e., slope of $\sim -0.6$).
We see that the computed feeding rate distribution shifts to the right $\sim 0.5~$dex per unit redshift,
indicating that the duty cycle of luminous AGNs increases with redshift.

\section{Conclusions}

We have shown that, baring implausible physical fine-tuning,
neither the central massive objects - SMBHs or NSCs - 
nor the stellar bulges can be regulated by blowing away the majority of gas that has already landed,
to explain the observed CMO-bulge relation.
This leaves us with only one viable option. That is, the ratio of feeding rate to the nuclear region
to that to the bulge is proportioned cosmologically. 

We test this scenario using high resolution, large-scale 
cosmological hydrodynamic simulations without AGN feedback.
Our analysis finds a proportionality, $\sim 0.1-0.3\%$,
between the feeding rate of very low angular momentum gas that can free-fall to the sub-parsec region 
to accrete to the SMBH and the star formation rate in the galaxy.
There is indication that this ratio
increases with increasing redshift to within a factor of $\sim 2$,
suggesting that the SMBH to bulge ratio is nearly redshift independent, 
with a modest increase with redshift.  
We predict that the duty cycle of luminous AGNs increases with redshift.
While SMBHs and bulges are found to coevolve on $\ge 30-150$Myr time scales,
there is indication that, on smaller time scales, the SMBH accretion and star formation may be less or not correlated,
which is likely due to variations of AGN activities on smaller time scales dictated by physics of accretion disc.

While our analysis disfavor internal regulation in terms of blowing gas away with the required proportionality,
``random" internal regulation by blowing some gas away without the said proportionality is not ruled out and in fact may be common,
manifested as galactic superwinds or AGN winds.
Nor do we disfavor feedback processes that control the overall amount of cold gas supply, termed "global feedback".
Global feedback reflects the collective effects of 
stellar evolution (supernovae, winds, etc) and SMBH accretion (winds, radio jets, etc) as well as
gravitational shock heating due to structure formation and photoionization heating, among others. 
They impact the thermodynamical state of the interstellar, circumgalactic and intergalactic medium.
We emphasize that, even if global feedback controls the overall cold gas supply 
and its temporal distribution on cosmological time scales,
it is not responsible for the proportional growth of SMBHs and galaxies.

An implication is that the distinction between forming a NSC or SMBH
may hinge on the existence of a massive enough initial black hole seed.
Thus, the demarcation bulge mass of $\mbg_{0}=(3-4)\times 10^{10}\msun$ is suggestive 
that only the progenitors of the massive enough galaxies have formed massive black hole seeds at some high redshift,
with less massive galaxies seeded by NSCs or neither. 
Subsequently, those with initial massive black hole seeds are able to accrete the infallen gas and grow to SMBHs over time, 
whereas those without massive black hole seeds turn the infallen gas in the nuclear regions into stars to grow the NSCs.
Let us suppose that 
CMOs of initial mass $M_{\rm CMO,init}$ created at some high redshift in dwarf galaxies
have migrated to the centers of larger galaxies to serve as central seeds.
The rationed gas supply would then yield  
final $M_{\rm CMO}/\mbg=(\alpha \mbg +M_{\rm CMO,init})/(\mbg + M_{\rm CMO, init})$.
Thus, for those galaxies lacking significant, subsequent growth of the CMO,
i.e., $\alpha \mbg$ is not much greater than $M_{\rm CMO,init}$,
the CMO-bulge mass scaling relation will be sublinear, which may explain the observed
shallower scaling relation between NSCs and bulges at the low end of bulge mass
\citep[e.g.,][]{2012Erwin, 2012Leigh, 2013Scott, 2014denBrok}.
Galaxies with a massive initial black hole seed may form a NSC as well,
consistent with observations \citep[e.g.,][]{2008Seth,2008GonzalezDelgado},
although the stellar component in the vicinity of an SMBH may be altered by subsequent, additional processes,
such inspiral of another SMBH \citep[e.g.,][]{2002Milosavljevic}.

This study is related to \citet[][]{2006Escala,2007Escala}, who studied gas accretion processes
surrounding the SMBH; we explicitly avoid detailed accretion physics by focusing
on the amount of mass that enters the ``feeding" zone to the SMBH. 
This work reaches conclusions similar to that of \citet[][]{2015AnglesAlcazar}
with respect to the $\mbh-\mbg$ ratio, with a contrasting difference on the role of feedback.
While \citet[][]{2015AnglesAlcazar} requires that only a small fraction of the gas at subparsec scales is actually
accreted by the SMBH, with the rest lost to winds and outflows,
we suggest that the gas disc beyond the Toomre unstable radius is instead consumed by star formation,
without requiring blowing away most of the gas by the SMBH.

\vskip 1cm

I am indebted to an anonymous referee for the most detailed, cogent, critical yet civilized reports,
which have immensely helped improve the presentation and clarify numerous issues.
I thank Dr. Guangtun Zhu for very helpful discussion.
This work is supported in part by grant NASA NNX11AI23G.


\begin{thebibliography}{44}
\expandafter\ifx\csname natexlab\endcsname\relax\def\natexlab#1{#1}\fi

\bibitem[{{Adams} {et~al.}(2001){Adams}, {Graff}, \& {Richstone}}]{2001Adams}
{Adams}, F.~C., {Graff}, D.~S., \& {Richstone}, D.~O. 2001, \apjl, 551, L31

\bibitem[{{Aird} {et~al.}(2012){Aird}, {Coil}, {Moustakas}, {Blanton},
  {Burles}, {Cool}, {Eisenstein}, {Smith}, {Wong}, \& {Zhu}}]{2012Aird}
{Aird}, J., {Coil}, A.~L., {Moustakas}, J., {Blanton}, M.~R., {Burles}, S.~M.,
  {Cool}, R.~J., {Eisenstein}, D.~J., {Smith}, M.~S.~M., {Wong}, K.~C., \&
  {Zhu}, G. 2012, \apj, 746, 90

\bibitem[{{Angl{\'e}s-Alc{\'a}zar} {et~al.}(2015){Angl{\'e}s-Alc{\'a}zar},
  {{\"O}zel}, {Dav{\'e}}, {Katz}, {Kollmeier}, \&
  {Oppenheimer}}]{2015AnglesAlcazar}
{Angl{\'e}s-Alc{\'a}zar}, D., {{\"O}zel}, F., {Dav{\'e}}, R., {Katz}, N.,
  {Kollmeier}, J.~A., \& {Oppenheimer}, B.~D. 2015, \apj, 800, 127

\bibitem[{{Cen}(2007)}]{2007Cen}
{Cen}, R. 2007, \apjl, 654, L37

\bibitem[{{Cen}(2014)}]{2014Cen}
---. 2014, \apj, 781, 38

\bibitem[{{Cen} {et~al.}(2005){Cen}, {Nagamine}, \& {Ostriker}}]{2005Cen}
{Cen}, R., {Nagamine}, K., \& {Ostriker}, J.~P. 2005, \apj, 635, 86

\bibitem[{{Cen} \& {Ostriker}(1992)}]{1992CenOstriker}
{Cen}, R., \& {Ostriker}, J.~P. 1992, \apjl, 399, L113

\bibitem[{{Colgate} {et~al.}(2003){Colgate}, {Cen}, {Li}, {Currier}, \&
  {Warren}}]{2003Colgate}
{Colgate}, S.~A., {Cen}, R., {Li}, H., {Currier}, N., \& {Warren}, M.~S. 2003,
  \apjl, 598, L7

\bibitem[{{C{\^o}t{\'e}} {et~al.}(2006){C{\^o}t{\'e}}, {Piatek}, {Ferrarese},
  {Jord{\'a}n}, {Merritt}, {Peng}, {Ha{\c s}egan}, {Blakeslee}, {Mei}, {West},
  {Milosavljevi{\'c}}, \& {Tonry}}]{2006Cote}
{C{\^o}t{\'e}}, P., {Piatek}, S., {Ferrarese}, L., {Jord{\'a}n}, A., {Merritt},
  D., {Peng}, E.~W., {Ha{\c s}egan}, M., {Blakeslee}, J.~P., {Mei}, S., {West},
  M.~J., {Milosavljevi{\'c}}, M., \& {Tonry}, J.~L. 2006, \apjs, 165, 57

\bibitem[{{Croton} {et~al.}(2006){Croton}, {Springel}, {White}, {De Lucia},
  {Frenk}, {Gao}, {Jenkins}, {Kauffmann}, {Navarro}, \& {Yoshida}}]{2006Croton}
{Croton}, D.~J., {Springel}, V., {White}, S.~D.~M., {De Lucia}, G., {Frenk},
  C.~S., {Gao}, L., {Jenkins}, A., {Kauffmann}, G., {Navarro}, J.~F., \&
  {Yoshida}, N. 2006, \mnras, 365, 11

\bibitem[{{den Brok} {et~al.}(2014){den Brok}, {Peletier}, {Seth}, {Balcells},
  {Dominguez}, {Graham}, {Carter}, {Erwin}, {Ferguson}, {Goudfrooij}, {Guzman},
  {Hoyos}, {Jogee}, {Lucey}, {Phillipps}, {Puzia}, {Valentijn}, {Verdoes
  Kleijn}, \& {Weinzirl}}]{2014denBrok}
{den Brok}, M., {Peletier}, R.~F., {Seth}, A., {Balcells}, M., {Dominguez}, L.,
  {Graham}, A.~W., {Carter}, D., {Erwin}, P., {Ferguson}, H.~C., {Goudfrooij},
  P., {Guzman}, R., {Hoyos}, C., {Jogee}, S., {Lucey}, J., {Phillipps}, S.,
  {Puzia}, T., {Valentijn}, E., {Verdoes Kleijn}, G., \& {Weinzirl}, T. 2014,
  ArXiv e-prints

\bibitem[{{Di Matteo} {et~al.}(2005){Di Matteo}, {Springel}, \&
  {Hernquist}}]{2005DiMatter}
{Di Matteo}, T., {Springel}, V., \& {Hernquist}, L. 2005, \nat, 433, 604

\bibitem[{Eisenstein \& Hut(1998)}]{1998Eisenstein}
Eisenstein, D.~J., \& Hut, P. 1998, ApJ, 498, 137

\bibitem[{{Erwin} \& {Gadotti}(2012)}]{2012Erwin}
{Erwin}, P., \& {Gadotti}, D.~A. 2012, Advances in Astronomy, 2012, 4

\bibitem[{{Escala}(2006)}]{2006Escala}
{Escala}, A. 2006, \apjl, 648, L13

\bibitem[{{Escala}(2007)}]{2007Escala}
---. 2007, \apj, 671, 1264

\bibitem[{{Faucher-Gigu{\`e}re} \& {Quataert}(2012)}]{2012FG}
{Faucher-Gigu{\`e}re}, C.-A., \& {Quataert}, E. 2012, \mnras, 425, 605

\bibitem[{{Ferrarese} {et~al.}(2006){Ferrarese}, {C{\^o}t{\'e}}, {Dalla
  Bont{\`a}}, {Peng}, {Merritt}, {Jord{\'a}n}, {Blakeslee}, {Ha{\c s}egan},
  {Mei}, {Piatek}, {Tonry}, \& {West}}]{2006Ferrarese}
{Ferrarese}, L., {C{\^o}t{\'e}}, P., {Dalla Bont{\`a}}, E., {Peng}, E.~W.,
  {Merritt}, D., {Jord{\'a}n}, A., {Blakeslee}, J.~P., {Ha{\c s}egan}, M.,
  {Mei}, S., {Piatek}, S., {Tonry}, J.~L., \& {West}, M.~J. 2006, \apjl, 644,
  L21

\bibitem[{{Ferrarese} \& {Merritt}(2000)}]{2000Ferrarese}
{Ferrarese}, L., \& {Merritt}, D. 2000, \apjl, 539, L9

\bibitem[{{Gebhardt} {et~al.}(2000){Gebhardt}, {Bender}, {Bower}, {Dressler},
  {Faber}, {Filippenko}, {Green}, {Grillmair}, {Ho}, {Kormendy}, {Lauer},
  {Magorrian}, {Pinkney}, {Richstone}, \& {Tremaine}}]{2000Gebhardt}
{Gebhardt}, K., {Bender}, R., {Bower}, G., {Dressler}, A., {Faber}, S.~M.,
  {Filippenko}, A.~V., {Green}, R., {Grillmair}, C., {Ho}, L.~C., {Kormendy},
  J., {Lauer}, T.~R., {Magorrian}, J., {Pinkney}, J., {Richstone}, D., \&
  {Tremaine}, S. 2000, \apjl, 539, L13

\bibitem[{{Gonz{\'a}lez Delgado} {et~al.}(2008){Gonz{\'a}lez Delgado},
  {P{\'e}rez}, {Cid Fernandes}, \& {Schmitt}}]{2008GonzalezDelgado}
{Gonz{\'a}lez Delgado}, R.~M., {P{\'e}rez}, E., {Cid Fernandes}, R., \&
  {Schmitt}, H. 2008, \aj, 135, 747

\bibitem[{{Goodman}(2003)}]{2003Goodman}
{Goodman}, J. 2003, \mnras, 339, 937

\bibitem[{{H{\"a}ring} \& {Rix}(2004)}]{2004Haring}
{H{\"a}ring}, N., \& {Rix}, H. 2004, \apjl, 604, L89

\bibitem[{{Hickox} {et~al.}(2014){Hickox}, {Mullaney}, {Alexander}, {Chen},
  {Civano}, {Goulding}, \& {Hainline}}]{2014Hickox}
{Hickox}, R.~C., {Mullaney}, J.~R., {Alexander}, D.~M., {Chen}, C.-T.~J.,
  {Civano}, F.~M., {Goulding}, A.~D., \& {Hainline}, K.~N. 2014, \apj, 782, 9

\bibitem[{{Hopkins} {et~al.}(2006){Hopkins}, {Hernquist}, {Cox}, {Di Matteo},
  {Robertson}, \& {Springel}}]{2006Hopkins}
{Hopkins}, P.~F., {Hernquist}, L., {Cox}, T.~J., {Di Matteo}, T., {Robertson},
  B., \& {Springel}, V. 2006, \apjs, 163, 1

\bibitem[{{Hopkins} \& {Quataert}(2010)}]{2010Hopkins}
{Hopkins}, P.~F., \& {Quataert}, E. 2010, \mnras, 407, 1529

\bibitem[{{Hopkins} \& {Quataert}(2011)}]{2011Hopkins}
---. 2011, \mnras, 415, 1027

\bibitem[{{Kauffmann} \& {Haehnelt}(2000)}]{2000Kauffmann}
{Kauffmann}, G., \& {Haehnelt}, M. 2000, \mnras, 311, 576

\bibitem[{{Komatsu} {et~al.}(2011){Komatsu}, {Smith}, {Dunkley}, {Bennett},
  {Gold}, {Hinshaw}, {Jarosik}, {Larson}, {Nolta}, {Page}, {Spergel},
  {Halpern}, {Hill}, {Kogut}, {Limon}, {Meyer}, {Odegard}, {Tucker}, {Weiland},
  {Wollack}, \& {Wright}}]{2011Komatsu}
{Komatsu}, E., {Smith}, K.~M., {Dunkley}, J., {Bennett}, C.~L., {Gold}, B.,
  {Hinshaw}, G., {Jarosik}, N., {Larson}, D., {Nolta}, M.~R., {Page}, L.,
  {Spergel}, D.~N., {Halpern}, M., {Hill}, R.~S., {Kogut}, A., {Limon}, M.,
  {Meyer}, S.~S., {Odegard}, N., {Tucker}, G.~S., {Weiland}, J.~L., {Wollack},
  E., \& {Wright}, E.~L. 2011, \apjs, 192, 18

\bibitem[{{Leigh} {et~al.}(2012){Leigh}, {B{\"o}ker}, \& {Knigge}}]{2012Leigh}
{Leigh}, N., {B{\"o}ker}, T., \& {Knigge}, C. 2012, \mnras, 424, 2130

\bibitem[{{Magorrian} {et~al.}(1998){Magorrian}, {Tremaine}, {Richstone},
  {Bender}, {Bower}, {Dressler}, {Faber}, {Gebhardt}, {Green}, {Grillmair},
  {Kormendy}, \& {Lauer}}]{1998Magorrian}
{Magorrian}, J., {Tremaine}, S., {Richstone}, D., {Bender}, R., {Bower}, G.,
  {Dressler}, A., {Faber}, S.~M., {Gebhardt}, K., {Green}, R., {Grillmair}, C.,
  {Kormendy}, J., \& {Lauer}, T. 1998, \aj, 115, 2285

\bibitem[{{Marconi} \& {Hunt}(2003)}]{2003Marconi}
{Marconi}, A., \& {Hunt}, L.~K. 2003, \apjl, 589, L21

\bibitem[{{Mestel}(1963)}]{1963Mestel}
{Mestel}, L. 1963, \mnras, 126, 553

\bibitem[{{Milosavljevi{\'c}} {et~al.}(2002){Milosavljevi{\'c}}, {Merritt},
  {Rest}, \& {van den Bosch}}]{2002Milosavljevic}
{Milosavljevi{\'c}}, M., {Merritt}, D., {Rest}, A., \& {van den Bosch}, F.~C.
  2002, \mnras, 331, L51

\bibitem[{{Murray} {et~al.}(2005){Murray}, {Quataert}, \&
  {Thompson}}]{2005Murray}
{Murray}, N., {Quataert}, E., \& {Thompson}, T.~A. 2005, \apj, 618, 569

\bibitem[{{Ostriker}(2000)}]{2000Ostriker}
{Ostriker}, J.~P. 2000, Physical Review Letters, 84, 5258

\bibitem[{{Ostriker} {et~al.}(2010){Ostriker}, {Choi}, {Ciotti}, {Novak}, \&
  {Proga}}]{2010Ostriker}
{Ostriker}, J.~P., {Choi}, E., {Ciotti}, L., {Novak}, G.~S., \& {Proga}, D.
  2010, \apj, 722, 642

\bibitem[{{Richstone} {et~al.}(1998){Richstone}, {Ajhar}, {Bender}, {Bower},
  {Dressler}, {Faber}, {Filippenko}, {Gebhardt}, {Green}, {Ho}, {Kormendy},
  {Lauer}, {Magorrian}, \& {Tremaine}}]{1998Richstone}
{Richstone}, D., {Ajhar}, E.~A., {Bender}, R., {Bower}, G., {Dressler}, A.,
  {Faber}, S.~M., {Filippenko}, A.~V., {Gebhardt}, K., {Green}, R., {Ho},
  L.~C., {Kormendy}, J., {Lauer}, T.~R., {Magorrian}, J., \& {Tremaine}, S.
  1998, \nat, 395, A14+

\bibitem[{{Scott} \& {Graham}(2013)}]{2013Scott}
{Scott}, N., \& {Graham}, A.~W. 2013, \apj, 763, 76

\bibitem[{{Seth} {et~al.}(2008){Seth}, {Ag{\"u}eros}, {Lee}, \&
  {Basu-Zych}}]{2008Seth}
{Seth}, A., {Ag{\"u}eros}, M., {Lee}, D., \& {Basu-Zych}, A. 2008, \apj, 678,
  116

\bibitem[{{Somerville} {et~al.}(2008){Somerville}, {Hopkins}, {Cox},
  {Robertson}, \& {Hernquist}}]{2008Somerville}
{Somerville}, R.~S., {Hopkins}, P.~F., {Cox}, T.~J., {Robertson}, B.~E., \&
  {Hernquist}, L. 2008, \mnras, 391, 481

\bibitem[{Tremaine {et~al.}(2002)Tremaine, Gebhardt, Bender, Bower, Dressler,
  Faber, Filippenko, Green, {et~al.}}]{2002Tremaine}
Tremaine, S., Gebhardt, K., Bender, R., Bower, G., Dressler, A., Faber, S.~M.,
  Filippenko, A.~V., Green, R., {et~al.} 2002, ApJ, 574, 740

\bibitem[{{Turner} {et~al.}(2012){Turner}, {C{\^o}t{\'e}}, {Ferrarese},
  {Jord{\'a}n}, {Blakeslee}, {Mei}, {Peng}, \& {West}}]{2012Turner}
{Turner}, M.~L., {C{\^o}t{\'e}}, P., {Ferrarese}, L., {Jord{\'a}n}, A.,
  {Blakeslee}, J.~P., {Mei}, S., {Peng}, E.~W., \& {West}, M.~J. 2012, \apjs,
  203, 5

\bibitem[{{Wehner} \& {Harris}(2006)}]{2006Wehner}
{Wehner}, E.~H., \& {Harris}, W.~E. 2006, \apjl, 644, L17

\end{thebibliography}

\end{document}